\documentclass[twocolumn,amsfont,amssymb,amsmath, showpacs,balancelastpage, nofootinbib]{revtex4-1}
\pdfoutput=1

\usepackage{aas_macros}
\usepackage[utf8]{inputenc}
\usepackage{geometry}
\usepackage{natbib}
\usepackage{graphicx}
\usepackage{amsmath}
\usepackage{xfrac}
\usepackage[colorlinks,linkcolor=red,citecolor=blue,urlcolor=blue]{hyperref}
\usepackage{cancel}
\usepackage{xcolor}
\usepackage{subfigure}
\usepackage{lipsum}
\newcommand{\ev}{\hat{\bf e}}
\newcommand{\nv}{\hat{\bf n}}
\newcommand{\dm}{{\cal D}}
\newcommand{\hcal}{{\cal H}}

\begin{document}

  \title{Linear anisotropies in dispersion-measure-based cosmological observables}
  \author{David Alonso$^1$}
  \affiliation{$^{1}$Department of Physics, University of Oxford, Denys Wilkinson Building, Keble Road, Oxford OX1 3RH, United Kingdom}

  \begin{abstract}
    We derive all contributions to the dispersion measure (DM) of electromagnetic pulses to linear order in cosmological perturbations, including both density fluctuations and relativistic effects. We then use this result to calculate the power spectrum of DM-based cosmological observables to linear order in perturbations. In particular we study two cases: maps of the dispersion measure from a set of localized sources (including the effects of source clustering), and fluctuations in the density of DM-selected sources. The impact of most relativistic effects is limited to large angular scales, and is negligible for all practical applications in the context of ongoing and envisaged observational programmes targetting fast radio bursts. We compare the leading contributions to DM-space clustering, including the effects of gravitational lensing, and find that the signal is dominated by the fluctuations in the free electron column density, rather than the local source clustering or lensing contributions. To compensate for the disappointing irrelevance of relativistic effects, we re-derive them in terms of the geodesic equation for massive particles in a perturbed Friedmann-Robertson-Walker metric.
  \end{abstract}

  \maketitle

  \section{Introduction}\label{sec:intro}
    Fast radio bursts (FRBs) \cite{1904.07947} are short ($\sim1-10\,{\rm ms}$) and bright ($\sim1\,{\rm Jy}$) radio pulses with a yet-unknown origin \cite{1810.05836}. Since their detection \cite{0709.4301}, interest in them has grown across all branches of astrophysics \cite{1604.01799}. The group velocity of these pulses changes with their frequency as they propagate through cold electron plasma, and the time delay between pulses at different frequencies is inversely proportional to their squared frequency ($\Delta t=(4.15\times10^{-3}\,{\rm s})\,{\rm DM}\,(1\,{\rm GHz}/\nu)^2$). The proportionality constant, the so-called ``dispersion measure'' (DM), is proportional to the column density of free electrons along the pulse's trajectory. Since FRBs are detected with DM values of $\gtrsim1000\,{\rm pc}\,{\rm cm}^{-3}$, while the Milky Way contribution to DM is $\lesssim 50\,{\rm pc}\,{\rm cm}^{-3}$ across most of the sky \citep{1610.09448}, FRB sources are most likely extragalactic. This is reinforced by their isotropic sky distribution \citep{1601.03547}, together with the identification of extragalactic persistent radio counterparts in long baseline interferometers \cite{1701.01098,1701.01099,1701.01100}. Our understanding of FRBs is likely to improve fast, with facilities like CHIME, UTMOST \cite{1601.02444} or HIRAX \cite{1607.02059} set up to provide catalogs containing thousands of FRBs.

    For these reasons, and in spite of the mystery surrounding their origin, FRBs have attracted the attention of cosmologists in recent years. Simple cosmological tests using DM as a distance measure \cite{1903.08175} are challenging, due to the scatter of the intergalactic component of the DM, and the degeneracy with the ionization fraction of the intergalactic medium (IGM) \cite{1812.11936}. DM can also be used to reconstruct the three-dimensional distribution of their sources \cite{1506.01704,1702.07085}, and the use of cross-correlations with optical surveys may allow us to estimate the redshift distribution of FRB sources and the clustering properties of the ionized IGM \cite{1912.09520}. Measurements of the DM can also be used as a probe of structure in the IGM, and help address the missing baryon problem \cite{1309.4451,1804.07291}, constrain the epoch of reionization \cite{1602.08130}, and break the optical depth degeneracy in observations of the kinematic Sunyaev-Zel'dovich effect \cite{1901.02418}, enabling a useful probe of structure growth \cite{1604.01382}. Other more ambitious proposals include using FRBs to test the equivalence principle \cite{1601.03636,2102.11554}, and even to constrain the level of primordial non-Gaussianity \cite{2007.04054}. 
    
    Although the prospect of using FRBs for precision cosmology lies in the future, given current uncertainties, it is important to have a complete quantitative description of the physical effects that DM-based observables are sensitive in order to best plan for that future. In this paper we provide a complete census of the gravitational relativistic effects that affect the propagation of electromagnetic pulses over cosmological distances at linear order in metric and density perturbations. We then quantify their impact on two DM-based cosmological observables based on anisotropies: the statistics of DM maps from a catalog of pulse-emitting sources, and the observed clustering of those sources when using DM as a radial distance proxy. The physical origin of the additional relativistic terms we will present is, for example, the modification of the pulse frequency (and hence its group velocity) due to gravitational redshifting and the integrated Sachs-Wolfe effect \cite{1967ApJ...147...73S}, as well as observational modifications to the inferred comoving distances and cosmic times assuming a background cosmological model. Most of these effects are only relevant on horizon-sized scales, and are largely unobservable \cite{1507.03550}. Nevertheless, it is relevant to quantify the expected amplitude of these General Relativistic contributions if e.g. tests of the equivalence principle from the large-scale distribution of dispersion measures are to be reliable \citep{1601.03636,2102.11554}. The results shown here are derived using methods similar to those presented in \cite{1105.5280,1105.5292} in the context of source number counts. 

    The paper is structured as follows. Section \ref{sec:dm} derives the expression for the dispersion measure of pulses propagating on a perturbed Friedmann-Robertson-Walker (FRW) backgound to linear order. This is then used in Section \ref{sec:dmcosmo} to derive expresions for the anisotropies of DM maps from pulse-emitting sources, and for the perturbation in the number counts of these sources using DM as a distance measure. We summarize our results and conclude in Section \ref{sec:conc}. Appendix \ref{app:geod} derives the solution to the geodesic equation for massive and massless particles in a perturbed FRW metric, offering an alternative derivation of the DM presented in Section \ref{sec:dm}, while \ref{app:cls} reviews a few standard results regarding the calculation of angular power spectra for projected cosmological anisotropies.
    
    Throughout the paper we will use natural units, where $\hbar=c=1$.

  \section{Time delay and dispersion measure}\label{sec:dm}
    \begin{figure}
      \centering
      \includegraphics[width=0.49\textwidth]{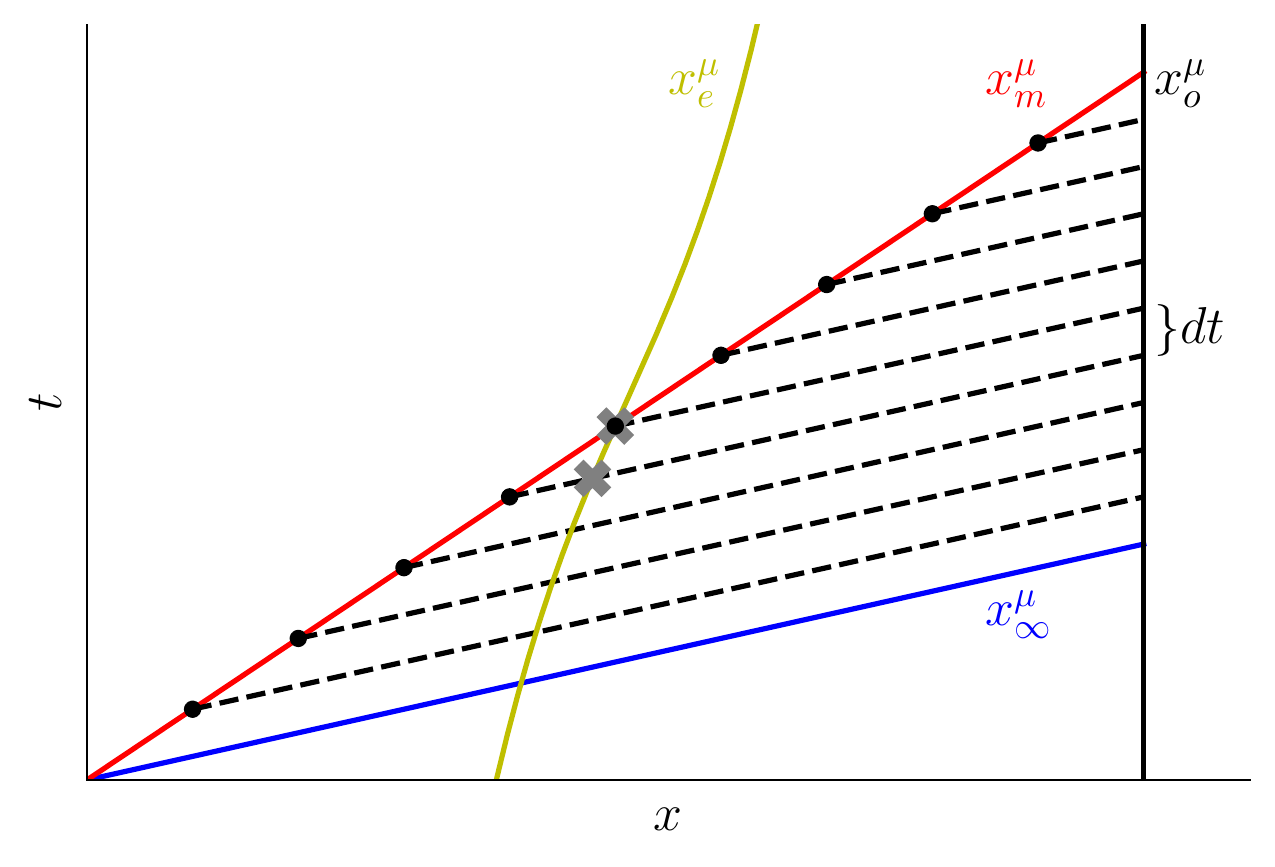}
      \caption{Spacetime diagram describing the calculation of the time delay for pulses propagating through the intergalactic plasma. The trajectories of the pulse, an infinite-frequency pulse, the final observer, and an observing station comoving with the plasma are shown in red, blue, black and yellow, and labelled $x_m^\mu$, $x_\infty^\mu$, $x_o$, and $x_e^\mu$ respectively. As the pulse arrives at the comoving station, an infinite-frequency pulse is emitted (black dashed lines) towards the final observer. The total time delay can be calculated as the sum of time intervals between the arrivals of these photons $dt$. This can be connected to the proper time measured by the comoving stations between the arrival of the pulse and the infinite-frequency pulse emitted by the previous observer (events marked with a cross in the figure).}\label{fig:xt}
    \end{figure}    
    An electromagnetic wave moving through cold plasma follows the dispersion relation:
    \begin{equation}
      \omega^2-|{\bf k}|^2=\omega_e^2,\hspace{12pt}\omega_e\equiv\frac{4\pi e^2n_e}{m_e},
    \end{equation}
    where $\omega$, ${\bf k}$, and $n_e$ are the wave's angular frequency, wave vector and electron number density, all measured in the rest frame of the electrons. In that frame, electromagnetic pulses propagate at a group velocity $v_g=\sqrt{1-\omega_e^2/\omega^2}$, which is equivalent to the motion of a massive particle with mass $\omega_e$ and energy $\omega$. For relevant frequencies and electron densities, the ratio $\omega_e/\omega$ is small \cite{2007.02886}, and thus we can approximate:
    \begin{equation}
      v_g\simeq1-\frac{1}{2}\frac{\omega_e^2}{\omega^2}
    \end{equation}

    Consider now a source emitting pulses at different frequencies. Our aim is to calculate the time delay $\Delta t$ measured by an observer between a pulse at observed frequency $\omega_o$, and a high-frequency pulse propagating at the speed of light. Following the analogy above, we will label these the ``massive'' and ``massless'' pulses respectively. To calculate this, consider a set of free-falling stations moving with the electron plasma between source and observer along the trajectory of the massive pulse. As the massive pulse passes through a given station, the station emits a massless pulse directed at the observer. The total time delay between the original massive and massless pulses is then the sum of the time intervals between the reception of the masless pulses emitted by each station. This setup is described in Figure \ref{fig:xt}. Let $d\tau$ be the proper time measured by one of the stations between the arrival of the previous station's massless pulse, and the massive one (events marked with crosses in the diagram). The corresponding time interval measured by the observer is $dt=(1+z)d\tau$, where $z$ is the station's redshift. The time interval measured by the station is the difference between the times taken by the massless and massive pulses to arrive from the preceding station, i.e.:
    \begin{equation}
      \frac{dt}{1+z}=d\tau=\frac{dl}{v_g}-dl,
    \end{equation}
    where $dl$ is the proper distance between the two stations. Writing the massless pulse's 4-momentum as $k^\mu=dx^\mu/d\lambda$, where $\lambda$ is an affine parameter, the proper length interval is $dl=(k^\mu u_{e,\mu})d\lambda$, where $u_{e,\mu}$ is the station's 4-velocity. Thus, to lowest order in $\omega_e/\omega$:
    \begin{equation}
      dt=\frac{(1+z)\omega_e^2}{2\omega^2}(k^\mu u_{e,\mu})d\lambda.
    \end{equation}
    Summing over all such time intervals, and relating $\omega$ to the observed frequency $\nu_o$ via $\omega=2\pi\nu_o(1+z)$, we obtain:
    \begin{equation}\label{eq:dm_def}
      \Delta t=\frac{e^2}{2\pi m_e\nu_o^2}\dm,
    \end{equation}
    where we have defined the \emph{dispersion measure} $\dm$ as
    \begin{equation}\label{eq:dm_general}
     \dm\equiv\int_{\lambda_e}^{\lambda_o}d\lambda \left.\left(n_e\frac{k^\mu u_{e,\mu}}{1+z}\right)\right|_\lambda,
    \end{equation}
    The subscript $\lambda$ is a reminder that the integrand is evaluated along the pulse's trajectory. Note that this is evaluated along a null geodesic (i.e. assuming the pulse propagates at the speed of light). Evaluating it along the massive pulse's trajectory would lead to higher-order corrections in $\omega_e/\omega$, which we neglect here.

    Let us consider now the propagation in a perturbed flat cosmological background. Considering only scalar perturbations, the perturbed FRW metric in the Newtonian gauge is:
    \begin{equation}\label{eq:metric}
      ds^2=a^2(\eta)\left[(1+2\psi)d\eta^2-(1-2\phi)\delta_{ij}dx^idx^j\right],
    \end{equation}
    where $\eta$ is the conformal time. Appendix \ref{sapp:geod.m0} presents the solution to the geodesic equation in this metric. Substituting Eqs. \ref{eq:eps_def}, \ref{eq:kmu_umu_m0} and \ref{eq:z_m0} into Eq. \ref{eq:dm_general} we obtain
    \begin{align}\label{eq:dm_pert}
      &\dm(\eta)=\int_{\eta}^{\eta_o}d\eta'\,a^2\,\bar{n}_e(\eta')\left[1+\tilde{\delta}_e\right]\\\nonumber
      &\tilde{\delta}_e(\eta)\equiv\delta_e+2\psi+\int_{\eta}^{\eta_o}d\eta'(\psi'+\phi'),
    \end{align}
    where we have decomposed $n_e$ into background $\bar{n}_e$ and perturbation $\bar{n}_e\delta_e$. Partial derivatives with respect to $\eta$ are denoted $\partial_\eta\xi\equiv\xi'$. This constitutes the main result of this paper. Appendix \ref{sapp:geod.mm} offers an alternative derivation of the same result from the geodesic equation of massive particles.

    Ignoring all perturbations we recover the usual expression for the background value
    \begin{equation}\label{eq:dm_bg}
      \bar{\dm}(z)\simeq\int_0^z\frac{dz'}{H(z')}(1+z')\bar{n}_{e,c}(z')
    \end{equation}
    where $\bar{n}_{e,c}\equiv a^3\bar{n}_e$ is the comoving electron density, and $H\equiv a^{-2}a'$ is the expansion rate. In what follows, we will model the background electron density as
    \begin{equation}\label{eq:ne_par}
      \bar{n}_{e,c}(z)=\frac{3H_0^2\Omega_b}{8\pi Gm_p}\frac{x_e(z)\,(1+x_H(z))}{2},
    \end{equation}
    where $x_e$ is the free electron fraction, $x_H$ is the hydrogen mass fraction, $\Omega_b$ is the baryon density fraction, and $m_p$ is the proton mass. For simplicity here we will use $x_H=0.75$ and $x_e=1$.

  \section{Dispersion measure-based cosmological observables}\label{sec:dmcosmo}
    In the context of the study of cosmological large-scale structure, measurements of $\dm$ from astrophysical sources can be used in two ways. First, since $\dm$ depends on the electron density and metric fluctuations integrated along the line of sight, $\dm$ can be used as a tracer of the projected large-scale structure in its own right (see e.g. \cite{1901.02418,2007.04054,2102.11554}). Secondly, assuming the background relation between $\dm$ and redshift (Eq. \ref{eq:dm_bg}), one can use the $\dm$ measured from a catalog of objects to reconstruct the three-dimensional large-scale structure (see e.g. \cite{1506.01704,1912.09520}). We discuss all the linear-order contributions to both observables in Sections \ref{ssec:dmcosmo.maps} and \ref{ssec:dmcosmo.counts}.

    \subsection{Dispersion measure as a structure tracer}\label{ssec:dmcosmo.maps}
      Given a set of pulse-emitting sources with known angular positions, it would be possible to make maps of the average local $\dm$ as a function sky position. Furthermore, radio interferometers are already able to localize FRB sources down to arc-second resolution \cite{1607.02059,2020Natur.581..391M}, and therefore we will be able to obtain redshifts for a large number of them in the future. These maps could then be made tomographically to potentially recover the three-dimensional distribution of free electrons. We study the properties of these maps here.

      Consider a set of pulse-emitting sources with known angular coordinates and redshift. The number of these sources in an solid angle $d\Omega$ and redshift interval $dz$ can be written as:
      \begin{equation}
        \frac{dN}{d\Omega dz}=\bar{n}_\Omega(z)[1+\tilde{\delta}_N],\hspace{6pt}\bar{n}_\Omega(z)\equiv\bar{n}_{s,c}(z)\frac{\chi^2(z)}{H(z)}
      \end{equation}
      where $\bar{n}_\Omega$ is source surface density, $\bar{n}_{s,c}(z)$ is the comoving density of sources, and $\chi(z)$ and $H(z)$ are the comoving distance and expansion rate, all evaluated at the time $\eta_*$ corresponding to the measured redshift in the background metric. $\tilde{\delta}_N$ collects all perturbations to the source number counts at linear order, which were presented in detail in \cite{1105.5280,1105.5292}.
      
      Likewise, the dispersion measure in Eq. \ref{eq:dm_pert} can be written as:
      \begin{equation}
        \dm=\bar{\dm}(z)\,[1+\delta_{\dm}],
      \end{equation}
      where $\bar{\dm}(z)$ is the background dispersion measure evaluated at $\eta_*$, and
      \begin{equation}
        \delta_{\dm}\equiv\frac{1}{\bar{\dm}(z)}\left[\int_{\eta_*}^{\eta_o}d\eta\,\bar{n}_e(\eta)\,a^2(\eta)\,\tilde{\delta}_e-\delta\eta\,a^2\,\bar{n}_e\right].
      \end{equation}
      Here $\delta\eta$ is the difference between the true conformal time and $\eta_*$. This can be found from Eq. \ref{eq:z_m0} to be \cite{1105.5292}:
      \begin{equation}
        \delta\eta=-\hcal^{-1}\left[\psi+\int_\eta^{\eta_o}(\psi'+\phi')-v_r\right],
      \end{equation}
      where $v_r$ is the radial peculiar velocity of the sources.

      Consider now a sample of sources selected in redshift with a selection function $w(z)$, and let $\dm_N(\nv)$ be the mean dispersion measure measured from these sources along direction $\nv$. The sky-averaged value is then given by
      \begin{equation}
        \bar{\dm}_N=\int dz\,p(z)\,\bar{\dm}(z),\hspace{12pt}
        p(z)\equiv \frac{w(z)\bar{n}_\Omega(z)}{\int dz' w(z')\,\bar{n}_\Omega(z')},
      \end{equation}
      where $p(z)$ is the normalized redshift distribution of the sample. The fluctuations around this mean are, to linear order, given by
      \begin{equation}\label{eq:dmmap_with_n}
       \delta\dm_N(\nv)=\int dz\,p(z)\,\bar{\dm}(z)\left[\delta_\dm+\left(1-\frac{\bar{\dm}_N}{\bar{\dm}(z)}\right)\tilde{\delta}_N\right].
      \end{equation}
      The second contribution is due to the clustering of the pulse sources, and appears at linear order because the quantity being traced ($\dm$) has a non-zero background value. This is in contrast with e.g. the second-order source clustering effects present in cosmic shear studies \citep{astro-ph/9712115,0910.3786}. In order to focus our attention on the relativistic effects inherent to $\dm$-based observables, in what follows we will ignore this second term for the most part, assuming that the evolution of $\bar{\dm}$ is negligible within the bin (i.e. assuming the prefactor of $\tilde{\delta}_N$ is negligible). As we will see, however, this second contribution is likely more relevant than any of these relativistic effects.

      In that case, expanding $\delta_{\dm}$ and $\delta\eta$, we obtain a final expression for the observed perturbation in $\dm_N$:
      \begin{widetext}
      \begin{equation}\label{eq:dm_map_total}
       \delta\dm_N(\nv)=\int_0^\infty\frac{dz}{H(z)}(1+z)\bar{n}_{e,c}P(>z)\left[\delta_e+(2-\rho)\psi+(1-\rho)\int_{\eta(z)}^{\eta_o}d\eta'(\psi'+\phi')+\rho\,v_r\right],
      \end{equation}
      \end{widetext}
      where we have defined the cumulative distribution $P(>z)$ and its slope $\rho$:
      \begin{equation}
        P(>z)\equiv\int_z^\infty dz'\,p(z'),\hspace{6pt} \rho\equiv\frac{1}{1+z}\frac{d\log P(>z)}{dz}.
      \end{equation}

      In order to compare the different contributions to the power spectrum of $\delta\dm_N$, this result can be Fourier-transformed and each perturbation connected with the primordial curvature perturbations via linear transfer functions \cite{1105.5292}. This procedure is outlined in Appendix \ref{app:cls}. The power spectrum between two $\delta\dm_N$ maps estimated from two galaxy samples $\alpha$ and $\beta$ is:
      \begin{equation}
        C^{\dm_N,(\alpha,\beta)}_\ell=4\pi \int \frac{dk}{k}\,{\cal P}_{\cal R}(k)\,\Delta^{\dm_N,\alpha}_\ell(k)\,\Delta^{\dm_N,\beta}_\ell(k),
      \end{equation}
      where
      \begin{widetext}
      \begin{align}\nonumber
        \Delta^{\dm_N,\alpha}_\ell(k)\equiv&\int_0^\infty d\chi\,(1+z)\bar{n}_{e,c}\,P^\alpha(>z)\left[b_eT_\delta(k,\eta_o-\chi)+(2-\rho)T_\psi(k,\eta_o-\chi)+\hcal(f_e-3)\frac{T_\theta(k,\eta_o-\chi)}{k^2}\right]j_\ell(k\chi)\\\nonumber
        &+\int_0^\infty d\chi\left(\int_\chi^\infty d\chi'(1+z')(1-\rho)\bar{n}_{e,c}P^\alpha(>z')\right)\,T_{\psi'+\phi'}(k,\eta_o-\chi)\,j_\ell(k\chi)\\\label{eq:cl_dmap}
        &-\int_0^\infty d\chi\,(1+z)\bar{n}_{e,c}\,P(>z)\,\rho(\chi)\,\chi\,T_\theta(k,\eta_o-\chi)\frac{j'_\ell(k\chi)}{k\chi}.
      \end{align}
      \end{widetext}
      Here $T_\xi(k,\eta)$ is the transfer function for perturbation $\xi$, $\theta=\nabla\cdot{\bf v}$ is the divergence of the peculiar velocity field, and we have assumed that the electron overdensity is a linearly biased tracer of the matter fluctuations $\delta$ in the synchronous gauge with bias $b_e$ \cite{0902.1084}. In this case, since $\delta_e$ is the linear overdensity in the Newtonian gauge:
      \begin{figure}
        \centering
        \includegraphics[width=0.49\textwidth]{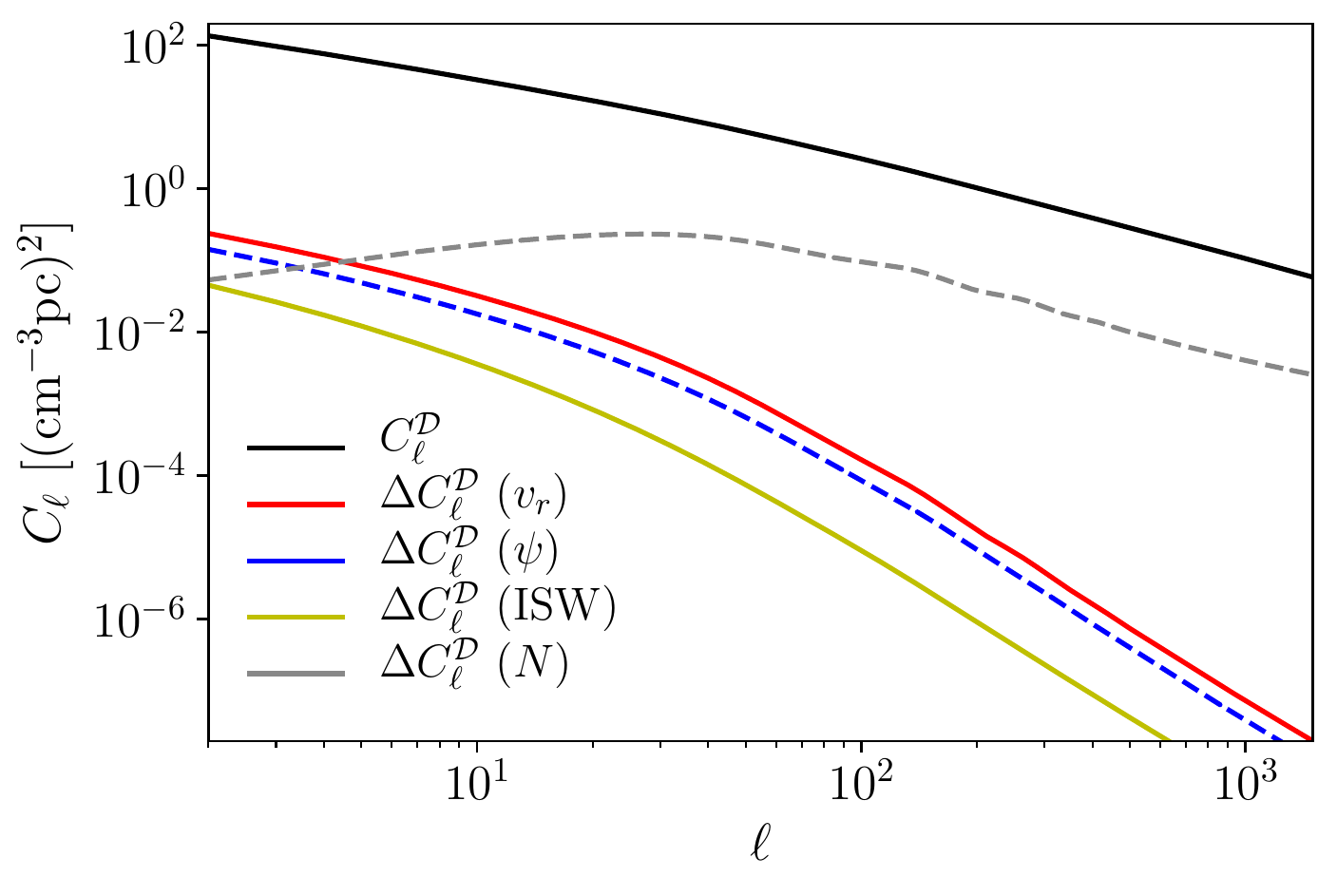}
        \caption{Angular power spectrum for a dispersion measure map from a sample of sources at redshift $z\sim1$ (black line). The different coloured lines show the contribution from different relativistic terms (including their cross-correlation with the electron density). Red, blue and yellow lines show the contribution from the two velocity terms, the gravitational potential term, and the ISW term respectively. The gray line shows the linear contribution from source clustering, which becomes significantly more relevant on small scales. Negative contributions are shown as dashed lines.}\label{fig:cl_dmap}
      \end{figure}    
      \begin{equation}
        \delta_e=b_e\delta+(f_e-3)\hcal\frac{\theta}{k^2},
      \end{equation}
      where $\hcal\equiv a'/a$, and we have defined the evolution bias
      \begin{equation}\label{eq:fe}
        f_e=\frac{d\log\bar{n}_{e,c}}{d\log a}.
      \end{equation}
      For our simple parametrization of $\bar{n}_{e,c}$ (Eq. \ref{eq:ne_par}) $f_e=0$.

      To simplify things further, we will assume the following transfer function relations, applicable at low redshifts in $\Lambda$CDM:
      \begin{align}
        &T_\theta=-\hcal\,f\,T_\delta,\hspace{12pt}
        T_\psi=T_\phi=-\frac{3\Omega_mH_0^2}{2ak^2}\,T_\delta,\\
        &T_{\phi'+\psi'}=-\frac{3\Omega_mH_0^2}{ak^2}\,\hcal\,(f-1)\,T_\delta,
      \end{align}
      where $f$ is the linear growth rate $f\equiv d\log\delta/d\log a$. All calculations were carried out using the Core Cosmology Library \cite{1812.05995}, using {\tt CAMB} to compute the linear matter power spectrum \cite{astro-ph/9911177}. We assume a flat $\Lambda$CDM cosmology with parameters $(\Omega_c,\Omega_b,h,\sigma_8,n_s)=(0.25, 0.05, 0.7, 0.8, 0.96)$. We also used Limber's approximation throughout \cite{1953ApJ...117..134L,astro-ph/0308260}:
      \begin{equation}
        j_\ell(x)\simeq \sqrt{\frac{\pi}{2\ell+1}}\delta^D(\ell+1/2-x).
      \end{equation}
      This is a valid approximation for the broad radial kernels considered here, and the inaccuracies it leads to at low $\ell$ do not change the qualitative results presented here.

      Figure \ref{fig:cl_dmap} shows the different contributions to the angular power spectrum for a dispersion measure map from a sample located at $z\simeq1$. Explicitly, we model the redshift distribution as a Gaussian centered at that redshift with width $\sigma_z=0.15$. For simplicity we assume unit bias $b_e=1$ for electrons, and $f_e=0$ (constant comoving electron density). The contributions from the two velocity terms, the potential term $\propto\psi$, and the ISW term in Eq. \ref{eq:cl_dmap}, including their correlation with the leading overdensity term are shown in blue, red and yellow respectively. We also include, in gray, the leading clustering contribution from Eq. \ref{eq:dmmap_with_n}, neglecting all relativistic and lensing terms in $\tilde{\delta}_N$, for a sample of sources with linear bias $b_s=1.5$. This contribution becomes significantly more relevant on small scales.

      The contribution from relativistic effects is largest on large scales, but negligibly small at all $\ell$, and completely undetectable even assuming a cosmic variance-limited map. Although it might be possible to beat this cosmic variance limit via multi-tracer methods \cite{0807.1770,0907.0707,1507.03550,1507.04605}, the prospects for this also seem unrealistic given the expected size of FRB samples ($<10^5$ sources). In turn, effects of source clustering are detectable at high significance (more than $20\sigma$ on $\ell<1000$ assuming cosmic variance uncertainties) and could become relevant for sufficiently dense FRB catalogs, especially in cross-correlation.

    \subsection{Dispersion measure as a distance proxy}\label{ssec:dmcosmo.counts}
      From the discussion above, it is evident that detecting the impact of the relativistic effects affecting the $\dm$-distance relation is challenging, and most likely unfeasible. However, the formalism developed thus far can be used to provide a complete census of the different effects that come into play at the linear level when using dispersion measure as a distance proxy to reconstruct the three-dimensional density distribution. We do so here.

      Consider a population of sources with known angular coordinates $\nv$ and dispersion measure $\dm$. Following a derivation similar to that of \cite{1105.5292}, the number of sources in a solid angle $d\Omega_o$ and dispersion measure interval $d\dm$ is related to the number density of sources via
      \begin{equation}\label{eq:dnddm_0}
        \frac{dN}{d\Omega_od\dm}=n_s\,\frac{dA_e}{d\Omega_o}\,(k^\mu u_{s,\mu})\,\frac{d\lambda}{d\dm},
      \end{equation}
      where $k^\mu$ is the wave vector of a light ray emitted by the sources, $u_s^\mu=a^{-1}(1-\psi,{\bf v})$ is the sources 4-velocity, $n_s$ is the source number density in their rest frame, $dA_e$ is the invariant area subtended by $d\Omega_o$, and $(k^\mu u_{s,\mu})d\lambda$ is the proper length interval covered by the ray in an interval $d\lambda$ of the affine parameter. Using the results found in Appendix \ref{app:geod} and \cite{1105.5292,1105.5280}, the different terms above are, to linear order:
      \begin{align}
        &n_s=\bar{n}_s(\eta)\,(1+\delta_n),\\
        &\frac{dA_e}{d\Omega}=a^2(\eta)\,\chi^2(\eta)\,(1-2\phi-2\kappa),\\\nonumber
        &(k^\mu u_{s,\mu})\frac{d\lambda}{d\dm}=(k^\mu u_{s,\mu})\frac{d\lambda}{d\eta}\frac{d\eta}{d\dm}\\
        &\hspace{52pt}=a(\eta)\,(1+\psi+v_r)\,\frac{d\eta}{d\dm},
      \end{align}
      where $\kappa$ is the lensing convergence
      \begin{equation}
        \kappa\equiv\nabla_{\nv}^2\int_0^\chi d\chi'\,\frac{\chi-\chi'}{\chi\chi'}\frac{\phi+\psi}{2},
      \end{equation}
      with $\nabla_{\nv}^2$ the angular Laplacian.

      $d\dm/d\eta$ can be calculated from Eq. \ref{eq:dm_pert}, and therefore
      \begin{equation}\label{eq:dnddm_1}
        \frac{dN}{d\Omega_od\dm}=\frac{a\bar{n}_s\chi^2}{\bar{n}_e}\left[1+\delta_n-\tilde{\delta}_e-2\phi-2\kappa+\psi+v_r\right].
      \end{equation}
      The prefactor in this equation is evaluated at conformal time $\eta$, which is different from the conformal time $\eta_{\dm}$ associated with the observed dispersion measure $\dm$ assuming no perturbations. The difference between both is, from Eq. \ref{eq:dm_pert}:
      \begin{equation}
        \delta\eta_{\dm}\equiv\eta-\eta_{\dm}=\frac{1}{a^2\bar{n}_e}\int_{\eta}^{\eta_o}d\eta'\,a^2\bar{n}_e\,\tilde{\delta}_e.
      \end{equation}
      Using this, the different components in the prefactor of Eq. \ref{eq:dnddm_1} are, to linear order:
      \begin{align}
        &a(\eta)=a(\eta_{\dm})\left[1+\hcal\,\delta\eta_{\dm}\right],\\
        &\bar{n}_x(\eta)=\bar{n}_x(\eta_{\dm})\left[1+(f_x-3)\hcal\,\delta\eta_{\dm}\right],\\
        &\chi^2(\eta)=\chi^2(\eta_{\dm})\left[1-\frac{2}{\chi}\left(\delta\eta_{\dm}-\int_\eta^{\eta_o}d\eta'\,(\psi+\phi)\right)\right],
      \end{align}
      where we have used \ref{eq:sol_i_m0} in the last line, and $f_x$ is the evolution bias for species $x$ as in Eq. \ref{eq:fe}.  Substituting everything back into Eq. \ref{eq:dnddm_1}, and expanding $\tilde{\delta}_e$, we obtain
      \begin{widetext}
      \begin{align}\nonumber
        \frac{dN}{d\Omega_od\dm}=\left(\frac{a\bar{n}_s\chi^2}{\bar{n}_e}\right)_{\dm}&\left[1+\delta_n-\delta_e+\left(1+f_s-f_e-\frac{2}{\hcal\chi}\right)
        \left(\frac{\hcal}{a^2\bar{n}_e}\int_\eta^{\eta_o}d\eta'a^2\bar{n}_e\left(\delta_e+2\psi+\int_{\eta'}^{\eta_o}d\eta''(\psi'+\phi')\right)\right)\right.\\\label{eq:dnddm_full}
        &\left.\,\,\,-2\phi-\psi-\int_\eta^{\eta_o}d\eta'(\psi'+\phi')-2\kappa+v_r+\frac{2}{\chi}\int_\eta^{\eta_o}d\eta'(\psi+\phi)\right].
      \end{align}
      \end{widetext}

      Eq. \ref{eq:dnddm_full} contains all linear-order contributions to the perturbation in the number of sources in ``dispersion measure space''. The terms involving $\delta_e$ are the $\dm$-space distortions introduced by \cite{1506.01704}, and can be interpreted as the perturbations to the observed comoving volume and cosmic time caused by using $\dm$ as a proxy for radial distance and time in the lightcone\footnote{Note that our expression for the so-called ``non-local'' term involving the integral of $\delta_e$ along the line of sight is different from that found by \cite{1506.01704}, which did not account for the change in $a^2\bar{n}_e$ as a function of distance.}. The term proportional to $\kappa$ is the usual lensing effect due to the apparent displacement of source positions. Note that we have not accounted for the effects of gravitational lensing on source flux, which would effectively modify the prefactor of this term by the slope of the source flux distribution leading to the standard source magnification effect. The specific value of this prefactor depends strongly on the selection function of the sample. Nevertheless, to provide an order-of-magnitude estimate of the correction due to lensing in comparison with the other contributions, we will simply keep this prefactor as is ($-2$). All other contributions are suppressed with respect to the $\propto\delta$ and $\propto\kappa$ terms by factors of $\hcal/k$ and $(\hcal/k)^2$, where $\hcal\sim2\times10^{-4}\,{\rm Mpc}^{-1}$ at $z\sim1$. This leads only to small modifications to the large-scale power spectrum that can be neglected, as discussed in the previous section (see also e.g. \cite{0907.0707,1505.07596,1507.03550}). Discarding these subdominant terms, and assuming a linear bias relation between $\delta_n$, $\delta_e$ and the synchronous-gauge matter overdensity $\delta$, the overdensity of $\dm$-selected sources in Eq. \ref{eq:dnddm_full} reduces to:
      \begin{figure}
        \centering
        \includegraphics[width=0.49\textwidth]{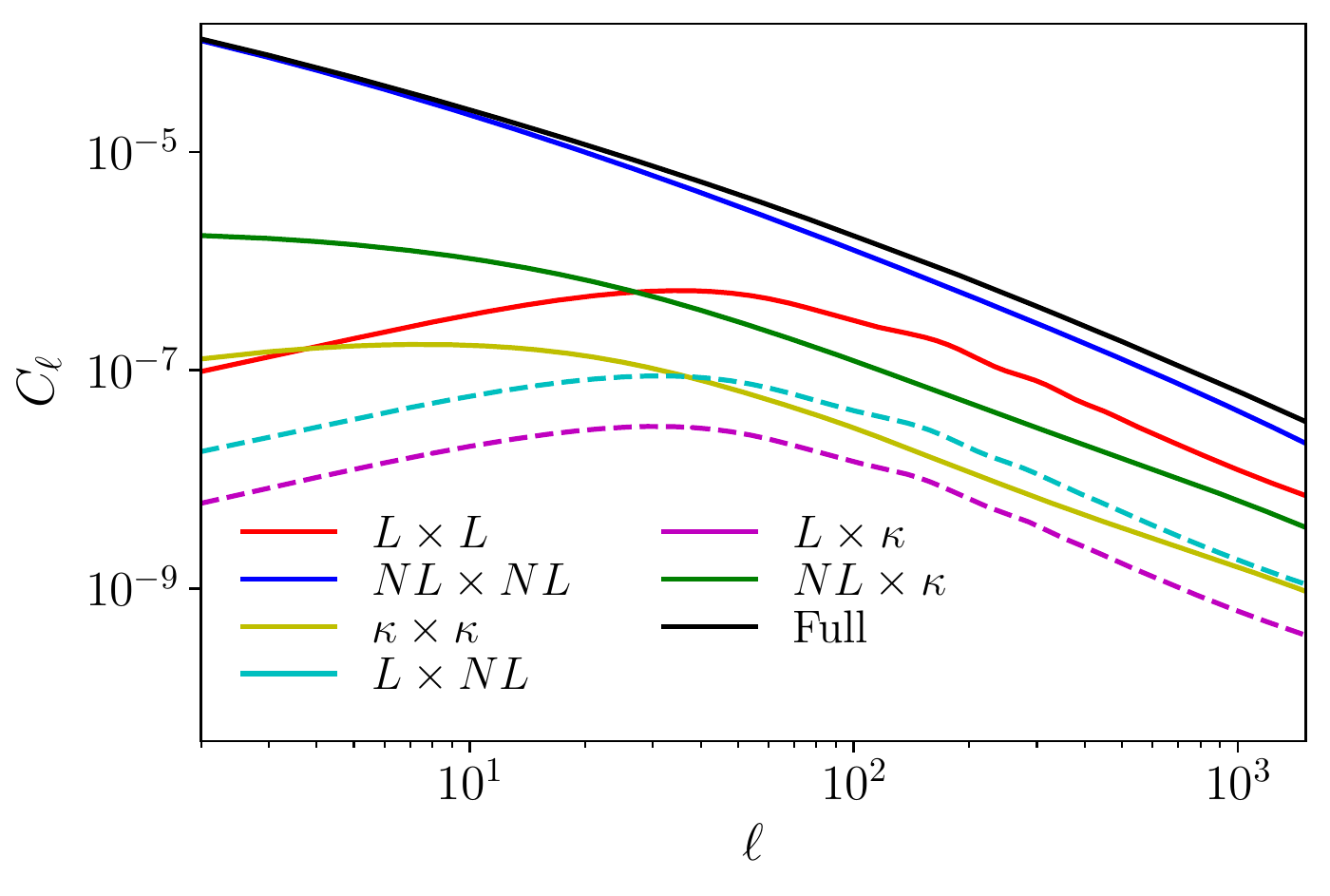}
        \caption{Angular power spectrum for the projected overdensity of DM-selected sources at $z\sim1$. The six coloured lines show the contributions from the auto- and cross-correlations of the ``local'', ``non-local'' and ``lensing'' terms in Eq. \ref{eq:dnddm_dom}, while the black line shows the total power spectrum. The signal is dominated on all scales by the non-local term, and the lensing contribution is subdominant to both the local and non-local components. Negative contributions are shown as dashed lines.}\label{fig:cl_ncdm}
      \end{figure}    
      \begin{align}\label{eq:dnddm_dom}
        \delta_{N\dm}=(b_n-b_e)\delta+{\cal A}(\chi)\frac{\hcal}{a^2\bar{n}_e}\int_0^\chi d\chi'a^2\bar{n}_e\,b_e\,\delta-2\kappa,
      \end{align}
      where we have defined
      \begin{equation}
        {\cal A}(\chi)\equiv 1+f_s-f_e-\frac{2}{\hcal\chi}.
      \end{equation}
      We will label the three contributions in Eq. \ref{eq:dnddm_dom} the ``local'' ($L$), ``non-local'' ($NL$) and ``lensing'' ($\kappa$) terms, in that order.

      The projected overdensity of sources in a dispersion-measure bin characterized by a DM distribution $p(\dm)$ is:
      \begin{align}\nonumber
        \Delta_{N\dm}(\nv)
        &=\int d\dm\,p(\dm)\,\delta_{N\dm}(\chi_{\dm}\nv,\eta_o-\chi_{\dm})\\
        &=\int dz\,\tilde{p}(z)\,\delta_{N\dm}(\chi(z)\nv,\eta_o-\chi(z)),
        \end{align}
      where in the second line we have translated $w(\dm)$ to redshift space for convenience via
      \begin{equation}
        \tilde{p}(z)=p(\bar{\dm}(z))\,\frac{d\bar{\dm}}{dz},
      \end{equation}
      with $\bar{\dm}(z)$ given by Eq. \ref{eq:dm_bg}. We can then make use of the procedure outlined in Appendix \ref{app:cls} to calculate the angular power spectrum of $\Delta_{N\dm}$.

      The result is shown in Figure \ref{fig:cl_ncdm}. The figure shows the contributions from the six auto- and cross-correlations of the $L$, $NL$ and $\kappa$ terms. The sources and electron density are modelled as in Section \ref{ssec:dmcosmo.maps} (Gaussian redshift distribution centered at $z=1$, $b_s=1.5$, $b_e=1$, $f_e=0$), and setting $f_s=0$ (constant comoving source density). As found in \cite{1506.01704}, the non-local contribution dominates the signal at all scales. The lensing contribution, new in this work, is smaller than both the local and non-local terms on all $\ell$s. As argued above, all other contributions in Eq. \ref{eq:dnddm_full} are only relevant on super-horizon scales, and can be safely neglected.
      
      This result is somewhat counter intuitive: although we have selected sources based on their radial distance, using $\dm$ as a proxy, their clustering signal is dominated by the distribution of the projected electron overdensity, rather than their intrinsic clustering. The main reason for this, is that the geometric perturbations to the volume in ``dispersion-measure space'' are significantly larger than the inhomogeneities in the source density. This is in contrast with the case of redshift-selected sources, where the impact of redshift-space distortions on the observed comoving volume are significantly milder. In addition to this, the partial cancellation betwee $\delta_s$ and $\delta_e$ in the local term also reduces its amplitude significantly if both quantities have biases $b_x\sim1$.

  \section{Conclusion}\label{sec:conc}
    We have presented a full derivation of all linear-order contributions to DM-based probes of cosmological anisotropies, including the impact of relativistic metric perturbations. These affect the dispersion measure directly by, for example, modifying the pulse frequencies and their group velocity via gravitational redshifting. We have then propagated these to two specific observables: the fluctuations of dispersion measure maps generated from a sample of localized sources, and the clustering of DM-selected sources. Most of these relativistic effects are suppressed with respect to density contributions by powers of $\hcal/k$, and thus are only potentially relevant on large, horizon-sized scales. A full description of these is nevertheless important if we are to use future FRB catalogs to constrain fundamental physics \citep{2007.04054,2102.11554}. Note that our calculation has not included other non-gravitational effects (e.g. scattering or absorption), as well as non-linearities and second-order lensing contributions. These may be more relevant than the relativistic effects presented here.

    As expected, we find that the impact of relativistic effects is small and mostly unobservable. Nevertheless, our formalism allows us to explore other contributions to the DM-induced anisotropies. We find that source clustering may be non-negligible when studying the fluctuations in dispersion measure maps, particularly in cross-correlation studies. We also find that the effects of gravitational lensing on the anisotropies of DM-selected sources are subdominant to that of geometric distortions in dispersion measure space, and that the latter likely dominate over the intrinsic source clustering signal. The use of dispersion measure as a tracer of large-scale anisotropies will thus provide a new window to study the properties of the inhomogeneous IGM over a wide range of scales and cosmic time, although doing so will also require a better understanding of the clustering of FRB sources.  
    \vspace{12pt}
 
  \section*{Acknowledgements}
    I would like to thank Deaglan Bartlett, Emilio Bellini, Harry Desmond, Pedro Ferreira, Mat Madhavacheril, and Andrina Nicola for useful discussions. DA is supported by the Science and Technology Facilities Council through an Ernest Rutherford Fellowship, grant reference ST/P004474. This publication arises from research funded by the John Fell Oxford University Press Research Fund.
  
  \bibliography{main}
  \appendix
  \onecolumngrid
  \section{Geodesic propagation of massive and massless particles}\label{app:geod}
    The temporal and spatial components of the geodesic equation for the perturbed FRW metric (Eq. \ref{eq:metric}) for a particle with 4-momentum $k^\mu\equiv dx^\mu/d\lambda$ read
    \begin{align}\label{eq:geod0_0}
      &\dot{k}^0+\left[\hcal+2\frac{d\psi}{d\eta}-\psi'\right]\,(k^0)^2+\left[\hcal(1-2\phi-2\psi)-\phi'\right]|{\bf k}|^2=0,\\\label{eq:geod0_1}
      &\dot{k}^i+\left[2\hcal-2\frac{d\phi}{d\eta}\right]k^ik^0+\partial_i\psi\,(k^0)^2+\partial_i\phi\,|{\bf k}|^2=0.
    \end{align}
    Here we denote derivatives with respect to the affine parameter $\dot{x}^\mu\equiv dx^\mu/d\lambda$, partial derivatives with respect to $\eta$ as $\alpha'\equiv\partial_\eta\alpha$, total derivatives with respect to $\eta$ as $d\alpha/d\eta\equiv \alpha'+k^i(\partial_i\alpha)/k^0$, and $\hcal\equiv a'/a$. These equations are complemented by the constraint on the norm of the 4-momentum which, for a particle of mass $m$, reads
    \begin{equation}\label{eq:mass}
      a^2\left[(1+2\psi)(k^0)^2-(1-2\phi)|{\bf k}|^2\right]=m^2.
    \end{equation}

    As in \cite{1105.5292}, we make the following change of variables: given a comoving observer with 4-velocity $u^\mu=a^{-1}(1-\psi)\delta^\mu_0$, we define the comoving energy $\epsilon$ as $k^\mu u_\mu\equiv \epsilon/a$. We then write the spatial components of the 4-momentum in terms of their modulus and direction vector $\ev$. Using Eq. \ref{eq:mass}, $\ev$ and $\epsilon$ are related to the components of $k^\mu\equiv\dot{x^\mu}$ via
    \begin{align}\label{eq:eps_def}
      \frac{d\eta}{d\lambda}=k^0=a^{-2}\epsilon(1-\psi),\hspace{12pt}
      \frac{d{\bf x}}{d\eta}=\frac{{\bf k}}{k^0}=(1+\psi+\phi)\sqrt{1-\gamma^{-2}}\,\ev,
    \end{align}
    where we have defined the Lorentz factor $\gamma\equiv \epsilon/(ma)$. The massless limit is achieved for $\gamma\rightarrow\infty$.

    In terms of these variables, the geodesic equations read:
    \begin{align}
      &\frac{d\epsilon}{d\eta}=\epsilon\left[-\frac{d\psi}{d\eta}+\psi'+\phi'+\gamma^{-2}(\hcal-\phi')\right]\\
      &\frac{d\ev}{d\eta}=-\frac{1}{(1-\gamma^{-2})^{1/2}}\nabla_\perp\psi-(1-\gamma^{-2})^{1/2}\nabla_\perp\phi,
    \end{align}
    where we have defined the transverse gradient $\nabla_\perp\equiv\nabla-\ev\,(\ev\cdot\nabla)$. For simplicity, we will set all perturbations at the observer to zero, which only contribute to monopole and dipole terms ignored here.

    \subsection{Solution for massless particles}\label{sapp:geod.m0}
      Setting $\gamma^{-2}=0$, the equations above can be solved to first order in the perturbations to yield:
      \begin{align}\label{eq:sol_0_m0}
        &\frac{\epsilon}{\epsilon_o}=1-\psi-\int_\eta^{\eta_o}(\psi'+\phi')d\eta',\\\label{eq:sol_i_m0}
        &{\bf x}=-\ev_o\int_\eta^{\eta_o}d\eta'(1+\phi+\psi)-\int_\eta^{\eta_o}d\eta'(\eta'-\eta)\nabla_\perp(\phi+\psi),
      \end{align}
      The redshift $z$ for a source with 4-velocity $u_s^\mu=a^{-1}[1-\psi,{\bf v}]$ is defined as $1+z\equiv(k^\mu u_{s,\mu})/(k^\mu u_{o,\mu})$, where the observer's 4-velocity is  $u_o^\mu\equiv \delta^\mu_0$. Using Eqs. \ref{eq:sol_0_m0} and \ref{eq:sol_i_m0} we find, to first order in the perturbations:
      \begin{equation}\label{eq:kmu_umu_m0}
        k^\mu u_{s,\mu}=a^{-1}\,\epsilon\,(1+v_r),
      \end{equation}
      where $v_r\equiv\nv\cdot{\bf v}$ is the velocity along the line of sight $\nv\equiv-\ev_o$. The redshift is therefore
      \begin{equation}\label{eq:z_m0}
        1+z=\frac{1}{a(\eta)}\left[1-\psi-\int_{\eta_o}^\eta d\eta'(\psi'+\phi')+v_r\right].
      \end{equation}

    \subsection{Time delay for small-mass particles}\label{sapp:geod.mm}
      For massive particles, in the small-mass limit $\gamma^{-2}\ll1$ the spatial geodesic equation reads
      \begin{align}
        &\frac{d\ev}{d\eta}\simeq-\nabla_\perp(\psi+\phi)-\frac{\gamma^{-2}}{2}\nabla_\perp(\psi-\phi).
      \end{align}
      Assuming $\psi=\phi$, this can be integrated to yield
      \begin{equation}\label{eq:sol_i_mm}
        {\bf x}=-\ev_o\int_\eta^{\eta_o}d\eta'(1+\phi+\psi)\left(1-\frac{\gamma^{-2}}{2}\right)-\int_\eta^{\eta_o}d\eta'\nabla_\perp(\phi+\psi)\int_\eta^{\eta'}d\eta''\left(1-\frac{\gamma^{-2}}{2}\right),
      \end{equation}

      Consider now two particles, one massless and one massive, emitted simultaneously by a source at coordinates $(\eta_e,{\bf x}_e)$, and reaching the observer at spatial coordinate ${\bf x}_o=0$ and times $\eta_{o,0}$ and $\eta_{o,m}$ respectively. From Eq. \ref{eq:sol_i_mm}, the radial coordinates of both particles at time $\eta$ are separated by
      \begin{equation}
        \chi_m(\eta)=\chi_0(\eta)+\int_{\eta_e}^\eta d\eta'\frac{\gamma^{-2}}{2}.
      \end{equation}
      A given perturbation $\xi$ evaluated along the massive geodesic is therefore related to the same quantity along the massless trajectory via:
      \begin{equation}\label{eq:geod_trans}
        \xi(\eta,\chi_m(\eta))=\xi(\eta,\chi_0(\eta))+\partial_\chi\xi\,\int_{\eta_e}^\eta d\eta'\frac{\gamma^{-2}}{2}=\xi(\eta,\chi_0(\eta))+\left(\xi'-\frac{d\xi}{d\eta}\right)\,\int_{\eta_e}^\eta d\eta'\frac{\gamma^{-2}}{2}.
      \end{equation}

      Evaluating Eq. \ref{eq:sol_i_mm} at $\eta=\eta_e$ for the massless and massive particles, subtracting them, projecting the result along $\ev_o$, and using Eq. \ref{eq:geod_trans} to relate $\phi$ and $\psi$ along both geodesics, we obtain:
      \begin{equation}
        0 = \eta_{o,m}-\eta_{o,0}-\int_{\eta_e}^{\eta_o}d\eta\frac{\gamma^{-2}}{2}\left(1-\int_\eta^{\eta_o}d\eta'(\psi'+\phi')\right).
      \end{equation}
      Writing $\gamma^{-2}$ in terms of $\epsilon$, and using Eq. \ref{eq:sol_0_m0}, we can write the time delay as
      \begin{equation}
        \Delta t=\int_{\eta_e}^{\eta_o}d\eta\frac{m^2a^2}{\epsilon_o^2}\left(1+2\psi+\int_\eta^{\eta_o}(\psi'+\phi')\right).
      \end{equation}
      This matches the expression found in Eqs. \ref{eq:dm_def} and \ref{eq:dm_pert} for $m=\omega_e$ and $\epsilon_o=2\pi\nu_o$.

    \section{Angular power spectra}\label{app:cls}
      Let $m(\nv)$ be a map of a projected quantity of the form:
      \begin{equation}\label{eq:map_gen}
        m(\nv)=\int d\chi \sum_\alpha {\cal O}^\alpha(\chi\nv, \eta(\chi)),
      \end{equation}
      where $\eta(\chi)=\eta_o-\chi$ is the conformal time at comoving distance $\chi$ in the background, and ${\cal O}^\alpha({\bf x},\eta)$ a set of linear perturbations. Let $q_{\ell m}$ be the spherical harmonic coefficients of a given function $q(\nv)$. In general, the harmonic coefficients of ${\cal O}^\alpha(\chi\nv,\eta)$ can be connected with the primordial curvature perturbations in Fourier space ${\cal R}({\bf k})$ via:
      \begin{equation}
        {\cal O}^\alpha_{\ell m}(\chi)=\int \frac{dk}{2\pi^2}k^2f^\alpha_\ell(k,\chi)i^\ell\int d\nv_k\,Y^*_{\ell m}(\nv_k){\cal R}({\bf k}),
      \end{equation}
      where $\nv_k$ is the unit vector in the direction of ${\bf k}$, $Y_{\ell m}(\nv)$ are the spherical harmonic functions, and the form of $f_\ell(k,\chi)$ depends on the contents of ${\cal O}^\alpha$. To find $f^\alpha_\ell$ for a given ${\cal O}^\alpha$, we express all perturbations in terms of their Fourier-space coefficients, connect them with ${\cal R}({\bf k})$, and use the plane wave expansion. The 5 most common forms for ${\cal O}^\alpha$ are:
      \begin{enumerate}
        \item Simple terms of the form ${\cal O}^\alpha=A(\eta)\,\xi(\chi\nv,\eta)$, where $\xi({\bf x},\eta)$ is a density or scalar metric perturbation. In this case, $f^\alpha_\ell(k,\chi)$ reads:
        \begin{equation}
          f_\ell^\alpha(k,\chi)=A(\eta_o-\chi)\,T_\xi(k,\eta_o-\chi)j_\ell(k\chi),
        \end{equation}
        where $j_\ell(x)$ is the order-$\ell$ spherical Bessel function, and $T_\xi$ is the transfer function relating $\xi$ and ${\cal R}$ ($\xi({\bf k},\eta)=T_\xi(k,\eta)\,{\cal R}({\bf k})$).
        \item Doppler terms of the form ${\cal O}^\alpha=A(\eta)\,v_r({\bf x},\eta)$, where $v_r$ is the radial peculiar velocity. In this case:
        \begin{equation}
          f_\ell^\alpha(k,\chi)=-\chi\,A(\eta_o-\chi)\,T_\theta(k,\eta_o-\chi)\frac{j'_\ell(k\chi)}{k\chi},
        \end{equation}
        where $\theta=\nabla\cdot{\bf v}$ is the divergence of the velocity field.
        \item Redshift-space distortion terms of the form ${\cal O}^\alpha=A(\eta)\partial_\chi v_r$. In this case:
        \begin{equation}
          f_\ell^\alpha(k,\chi)=-A(\eta_o-\chi)\,T_\theta(k,\eta_o-\chi)\,j''_\ell(k\chi).
        \end{equation}
        \item Integrated terms of the form ${\cal O}^\alpha=A(\eta)\int_0^\chi d\chi'\,B(\chi)\,\xi(\chi'\nv,\eta')$. In this case:
        \begin{equation}
          f_\ell^\alpha(k,\chi)=A(\eta_o-\chi)\,\int_0^\chi d\chi'\,B(\chi')\,T_\xi(k,\eta_o-\chi')\,j_\ell(k\chi').
        \end{equation}
        When further integrated over $\chi$, these terms can be simplified by swapping the order of the integrals over $\chi$ and $\chi'$, to yield terms of the form:
        \begin{equation}
          \int_0^\infty d\chi\,\left(\int_\chi^\infty d\chi'\,A(\eta_o-\chi')\right)B(\chi)\,T_\xi(k,\eta_o-\chi)\,j_\ell(k\chi).
        \end{equation}
      \end{enumerate}

      Then, using the definition of the primordial power spectrum
      \begin{equation}
        \langle {\cal R}({\bf k})\,{\cal R}^*({\bf k}')\rangle\equiv(2\pi)^3\delta^D({\bf k}-{\bf k}') \frac{2\pi^2}{k^3}{\cal P}_{\cal R}(k),
      \end{equation}
      the power spectrum of the map in Eq. \ref{eq:map_gen} is given by:
      \begin{equation}
        C_\ell=4\pi\sum_{\alpha\beta}\int \frac{dk}{k}\int d\chi\int d\chi'\,{\cal P}_{\cal R}(k)f^\alpha_\ell(k,\chi) \,f^\beta_\ell(k,\chi').
      \end{equation}

\end{document}